\documentstyle[aps,epsfig]{revtex}                         

\draft
\begin{document}
\twocolumn[\hsize\textwidth\columnwidth\hsize
     \csname @twocolumnfalse\endcsname



\title{Determining surface magnetization and local magnetic
       moments with atomic scale resolution.
       }

\author{W. A. Hofer and A. J. Fisher}
\address{
         Department of Physics and Astronomy,
         University College, Gower Street, London WC1E 6BT, UK}

\maketitle

\begin{abstract}
We propose a method to determine the direction of surface
magnetization and local magnetic moments on the atomic scale. The
method comprises high resolution scanning tunneling microscope
experiments in conjunction with first principles simulations of
the tunneling current. The potential of the method is demonstrated
on a model system, antiferromagnetic Mn overlayers on W(110). We
expect that it will ultimately allow to study the detailed changes
of magnetic surface structures in the vicinity of dopants or
impurities.
\end{abstract}

\pacs{72.25.-b,73.40.Jn,75.25.+z}

\vskip2pc]

Until very recently magnetic properties of thin metal films or
surfaces could be determined only by x-ray magnetic dichroism
experiments \cite{stoehr95,weller95,goering01}. In these
experiments x-rays of circular polarization, originating from
synchrotron radiation, are used to probe the magnetic properties.
With the help of second order perturbation theory the intensity of
the adsorbed radiation can be related to magnetic moments $\mu$
and the direction of spin polarization $\vec{M}$ due to crystal
anisotropies \cite{bruno89}. However, the method suffers from two
serious weaknesses, which limit its applicability: (i) The local
resolution obtained with synchrotron radiation is far too low for
any atomic scale analysis and in the range of 50nm
\cite{stoehr99}. (ii) The intensity of the x-ray beams is high
enough to lead to substantial energy dissipation. Since magnetic
properties are very sensitive to temperature changes, the
measurement of groundstate properties is problematic.

A solution to these problems could come from a different method,
one that does not affect the electronic states of a magnetic layer
in any substantial way. The spin polarized (SP) scanning tunneling
microscope (STM) does provide just such a method. In measurements
with an iron coated tungsten tip on single antiferromagnetic Mn
layers it has been demonstrated that the local magnetic moment can
be resolved on the atomic scale by STM scans \cite{heinze00}. The
theoretical simulation of these scans revealed that such a scan is
very sensitive to the chemical nature of the STM tip apex, which
allows one to identify the STM tip from the corrugation height of
the surface \cite{hofer01a}. However, these results were obtained
under the condition of ferromagnetic ordering of sample and tip
states, the direction of surface magnetization was therefore
imposed from the outset.

In this Letter we propose an extension of the method to account
for truly general orientations of the magnetic axis. We want to
show, how the orientation of the magnetization vector $\vec{M}$ of
sample and tip changes the tunneling current and the corrugation
height measured on a surface. This in turn allows one to determine
the magnetic axis and the local moments from STM scans and first
principles simulations. Since antiferrromagnetic surface ordering
on the atomic scale has already been observed \cite{heinze00}, it
is justified to assume that magnetic surface properties are not
decisively modified by an STM tip. And as the experiments are
suitable to reveal electronic properties on an atomic scale, they
also remove the resolution problem one is confronted with in
dichroism measurements.

Let us consider the situation in a tunnelling junction between a
crystal surface and an STM tip in real space. Magnetic anisotropy
in a crystal breaks the rotational symmetry of electron spins. The
spin states in this case are projected onto the crystal's magnetic
axis. We assume in the following that this symmetry breaking
occurs in the two separate systems which form our tunnelling
junction. Depending on the orientation of the magnetic axes two
limiting cases have to be distinguished. The magnetic axis of
sample and tip are either parallel or antiparallel. In the first
case we have to sum up all electrons tunnelling from spin-up
states of the sample ($n_S^{\uparrow}$) to spin-up states of the
tip ($n_T^{\uparrow}$). This ferromagnetic ordering is described
by the following transitions:

\begin{equation}
{\bf M}_S \, \uparrow \qquad \left\{ \begin{array}{c}
  n_S^{\uparrow} \quad \longrightarrow \quad n_T^{\uparrow}\\
  n_S^{\downarrow} \quad \longrightarrow \quad n_T^{\downarrow}
\end{array} \right\}
\qquad \uparrow {\bf M}_T
\end{equation}

Here ${\bf M}_S$ and ${\bf M}_T$ describe the magnetic axes of
sample and tip, respectively. We denote the tunnelling current due
to ferromagnetic ordering by $I_{F}$. If the two vectors are
antiparallel the electrons tunnel from states ($n_S^{\uparrow}$)
into states ($n_T^{\downarrow}$) and vice versa. The
antiferromagnetic ordering is therefore described by:

\begin{equation}
{\bf M}_S \, \uparrow \qquad \left\{ \begin{array}{c}
  n_S^{\uparrow} \quad \longrightarrow \quad n_T^{\downarrow}\\
  n_S^{\downarrow} \quad \longrightarrow \quad n_T^{\uparrow}
\end{array} \right\}
\qquad \downarrow {\bf M}_T
\end{equation}

This setup yields the antiferromagnetic tunnelling current
$I_{A}$. Since the energy of the tunnelling electrons is very low,
and since the overlap of the sample and tip wavefunctions is
computed far outside the core region of surface atoms, spin-orbit
coupling can generally be neglected in the theoretical treatment.
Within density functional theory (DFT) \cite{kohn64,kohn65} the
tunnelling current is commonly described in terms of $\phi_M$, the
angle between the two magnetic axes, and $P_{S(T)}$, the
polarization of the sample (tip) surface:

\begin{equation}
I(\phi_M) = I_0 (1 + P_S P_T \cos \phi_M)
\end{equation}

\begin{equation}
\cos \phi_M = \frac{{\bf M}_S \cdot {\bf M}_T}{|{\bf M}_S| |{\bf
M}_T|}
\end{equation}

For constant tunnelling matrix elements and within a perturbation
approach the current $I_0$ and polarizations $P_{S(T)}$ are given
by:

\begin{equation}
I_0 \propto \frac{1}{2}\left(n_{S}^{\uparrow} +
n_{S}^{\downarrow}\right) \left(n_{T}^{\uparrow} +
n_{T}^{\downarrow}\right) \end{equation}

\begin{equation}
P_{S(T)} = \frac{n_{S(T)}^{\uparrow} -
n_{S(T)}^{\downarrow}}{n_{S(T)}^{\uparrow} +
n_{S(T)}^{\downarrow}}
\end{equation}

$I_0$ and $P_S P_T$ can be written in terms of the ferromagnetic
and antiferromagnetic currents:

\begin{eqnarray}
I_0 = \frac{1}{2}\left(I_F + I_A\right) \qquad
P_S P_T = \frac{I_F
- I_A}{I_F + I_A}
\end{eqnarray}

If the tunnelling matrix element is not constant, the current has
to be calculated numerically from the Bardeen integral over the
separation surface \cite{bardeen61,hofer00a}. In this case the
current contributions have to account for the spin orientation of
a given eigenstate. In DFT the energetic minimum for magnetic
crystals is reached by optimizing the distribution of spin-up
density $n^{\uparrow}$ and spin-down density $n^{\downarrow}$. The
currents for ferromagnetic and antiferromagnetic coupling  are
computed by calculating the transition matrix elements for the
spin polarized Kohn-Sham states of sample and tip:

\begin{equation}
I_{F} = I (n_S^{\uparrow} \longrightarrow n_T^{\uparrow}) + I
(n_S^{\downarrow} \longrightarrow n_T^{\downarrow})
\end{equation}

\begin{equation}
I_{A} = I (n_S^{\uparrow} \longrightarrow n_T^{\downarrow}) + I
(n_S^{\downarrow} \longrightarrow n_T^{\uparrow})
\end{equation}

Calculating the tunnel current for different angles $\phi_M$
requires then only to compute the linear combination of
ferromagnetic and antiferromagnetic currents multiplied by the
appropriate coefficients. From these three dimensional current
maps the constant current contours and the surface corrugations
can be extracted in a straightforward manner.

\begin{figure}
\begin{center}
\epsfxsize=1.0\hsize \epsfbox{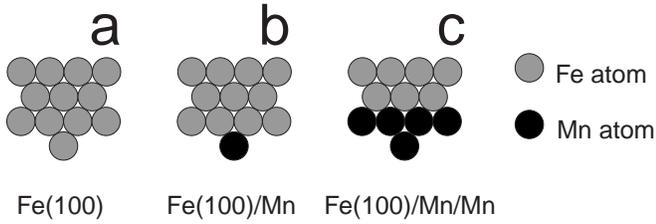}
\end{center}
\caption{STM tip models in our simulation. The tip is mimicked by
         a c(2$\times$2) Fe(100) free standing film with a single
         apex atom (a). Contamination of the tip by atoms of the
         crystal surface is accounted for by a single Mn apex
         atom (b), or an Mn monolayer and a Mn apex atom (c).
         }
\label{fig001}
\end{figure}

We have used a full potential method to compute the electronic
groundstate properties of model tips. The tip in the experiments
was a tungsten wire coated with several layers of iron
\cite{heinze00}. We mimic this tip by an ideal Fe(100) surface
with a single Fe atom in the apex (see Fig. \ref{fig001}).
Previous simulations of STM experiments revealed that the best
agreement between experiments and simulations is often obtained
with a tip model, covered by impurities of the sample surface
\cite{hofer00a}. Therefore we modified the clean Fe-tip by two
additional setups. In one case the apex atom was changed to Mn, in
the second also the surface layer consisted of Mn atoms (see Fig.
\ref{fig001}). On the technical side we note that the free
standing film consisted of five Fe layers and two additional
layers for the apex. 10 special k-points were used in the spin
polarized DFT calculations. Given the c(2$\times$2) unit cell this
amounts to 40 k-points for the elementary cell. The convergence in
the final iterations was better than 0.01 e/au$^3$. Since Mn and
Fe are both 3$d$ metals and relaxations are therefore rather
small, the STM simulations were based on the wavefunctions of bulk
truncated crystals. The details of the electronic structure
calculations are published in a separate paper \cite{hofer02a}.

In the Bardeen approach to tunnelling \cite{bardeen61} the
tunnelling current is computed by integrating the overlap of
sample and tip states over the separation surface. In our program,
which is described in detail elsewhere (bSCAN \cite{hofer00a}), we
integrate over a finite separation area and sum up all
contributions from the Kohn-Sham states of sample and tip
numerically. It has been shown previously that the STM current in
the experiments on W(110)Mn is about one to two orders of
magnitude higher than the current obtainable within a perturbation
approach and at a reasonable distance \cite{hofer01a}. Even though
we could not pin down this discrepancy, it seems to be most likely
due to current leakages of the STM circuit. In our implementation
of perturbation theory it is implicitly assumed that all
tunnelling current passes through a small area of the separation
surface, thus it cannot account for off center contributions in
the circuit. Therefore we computed the constant current contours
not for the actual values in the measurements (which were in the
range of 30nA \cite{heinze00}), but chose contours centered at
about 4.5 to 4.6 \AA \, above the W(110)Mn film. This distance is
at the lower limit of mechanical stability on metal surfaces
\cite{hofer01b}. The bias voltage in the measurements was - 3 mV,
we chose the same voltage for our simulations. In the figures we
display the constant current contours depending on two separate
parameters, important for every single STM experiment: (i) The
chemical nature of the tip surface. We report on simulations with
three different chemical compositions of the tip. (ii) The angle
of magnetization. We show simulations for selected angles $\phi_M$
between the magnetic axis of the crystal surface and the STM tip.

\begin{figure}
\begin{center}
\epsfxsize=1.0\hsize \epsfbox{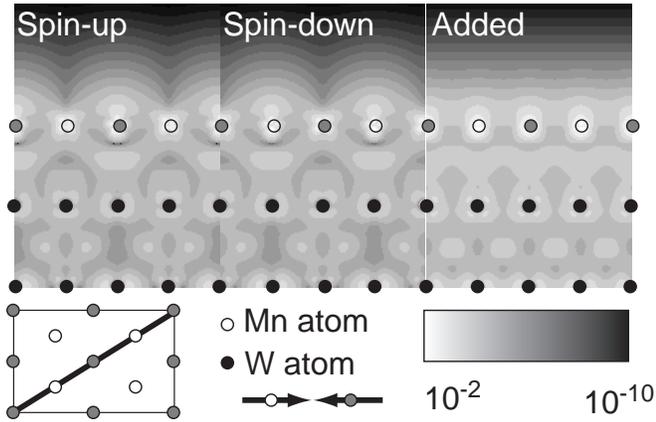}
\end{center}
\caption{Electronic structure of the W(110)Mn film. The plots show
the local density of states in a vertical plane cut through the
crystal (sketch at the bottom, left). The magnetic moment in the
surface atoms is either positive (empty circles), or negative
(grey circles). The LDOS above different atoms reveals these atoms
either as protrusions, or as depressions, depending on the
electron spin. The corrugation of the surface layers vanishes for
paramagnetic STM tips.
         }
\label{fig002}
\end{figure}

The electronic structure of the W(110)Mn surface in
antiferromagnetic ordering depends substantially on the spin
orientation of electron charge. For spin-up states the density
contours have a maximum at the position of Mn atoms with a
negative magnetic moment and their minimum at the position of
atoms with the opposite polarization (see Fig. \ref{fig002}). For
spin-down states the situation is reversed: atoms with positive
magnetic moments are now seen as protrusions. This indicates that
the surface has the highest corrugation for charge transport with
a high degree of polarization. In addition, it shows that the
surface is comparatively flat if measured by a paramagnetic tip
(see Fig. \ref{fig002}, right frame).

\begin{figure}
\begin{center}
\epsfxsize=1.0\hsize \epsfbox{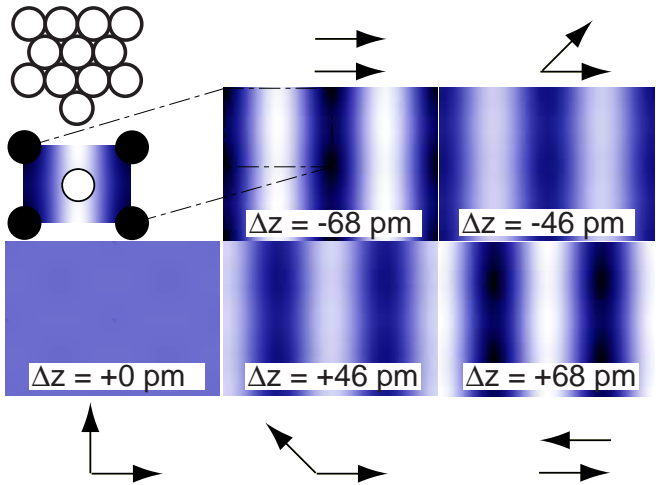}
\end{center}
\caption{Simulated STM scans for a Fe(100) tip with Fe apex. The
simulations show four unit cells, the orientation of the magnetic
axes for the individual scans is depicted by the black arrows. The
corrugation height $\Delta z$ is the difference in apparent height
between atoms of positive (full circles) and negative (empty
circle) magnetic moments. For perpendicular moments the surface
corrugation vanishes.
         }
\label{fig003}
\end{figure}

\begin{figure}
\begin{center}
\epsfxsize=1.0\hsize \epsfbox{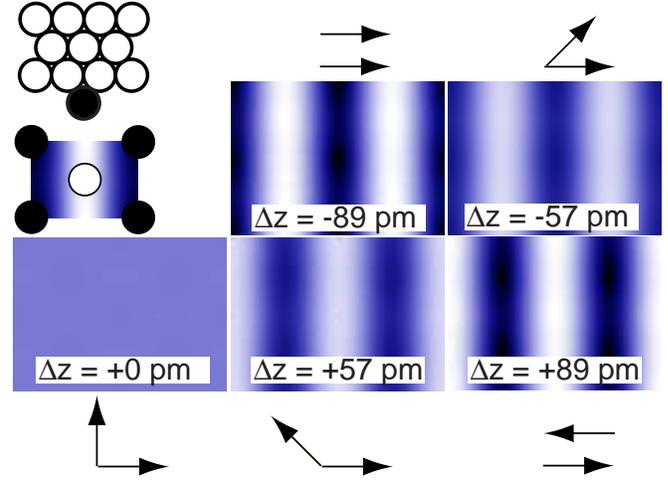}
\end{center}
\caption{Simulated STM scans for a Fe(100) tip with Mn apex. The
corrugation in this case is higher, the apparent height is greater
for surface atoms with a positive magnetic moment.
         }
\label{fig004}
\end{figure}

\begin{figure}
\begin{center}
\epsfxsize=1.0\hsize \epsfbox{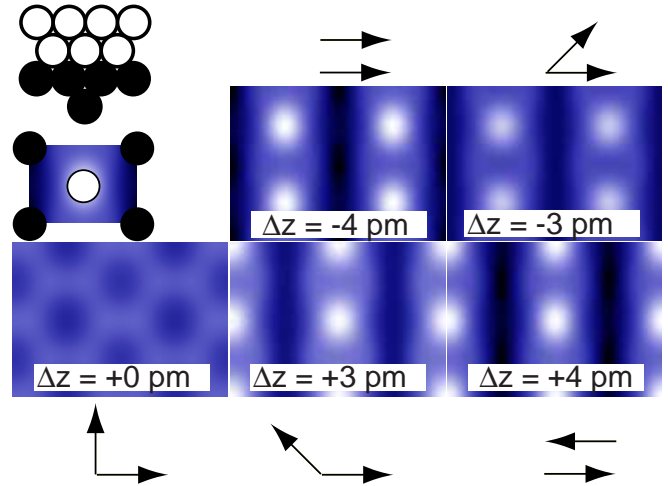}
\end{center}
\caption{Simulated STM scans for a Fe(100)Mn tip with Mn apex. The
magnetic contrast in this case nearly vanishes. The corrugation
height remains unchanged over a wide range of angles, contrary to
the simulations with clean or moderately contaminated STM tips.
         }
\label{fig005}
\end{figure}

Figs. \ref{fig003} to \ref{fig005} display the results of our STM
simulations with clean and Mn contaminated Fe model tips. We show
five angular settings, from $\phi_M = 0$ (ferromagnetic charge
transitions) to $\phi_M = 180^{\circ}$ (antiferromagnetic charge
transitions). In general the obtained magnetic contrast depends on
the angle $\phi_M$. It is a maximum for the limiting values of
$0^{\circ}$ and $180^{\circ}$, and it vanishes for $\phi_M =
90^{\circ}$. This angle denotes the case where tunnelling
transitions from the sample into the tip are essentially
unpolarized. We note two distinct features in our simulations:
\begin{itemize}
\item
The obtained maximum of magnetic contrast depends on the chemical
nature of the tip apex.
\end{itemize}
It is 68pm for a clean tip; 89pm for a tip contaminated with an Mn
atom; and it nearly vanishes for a tip contaminated by a surface
layer of Mn (4pm, Fig. \ref{fig005}). We may therefore conclude
that high resolution measurements with a suitable degree of
precision, measurements which are already possible at present,
allow to differentiate between different tips also in case of a
magnetic tunnel junction. For non-magnetic tunnel junctions, we
have already shown the influence of the tip in previous
publications \cite{hofer00a}. Comparing the simulations with the
experiments \cite{heinze00}, the most likely tip in the actual
experiments seems to be tip model (c) (see Fig. \ref{fig001}): in
this case the corrugation is within the range of experiments over
a wide range of angles (the range is about 2-4pm and it varies,
seemingly due to the lateral angle of surface polarization
\cite{bode01c}). This indicates that the STM tip during the
measurement is contaminated by at least a monolayer of surface
atoms; we would conclude from this feature that the tip is
generally very close to the surface and that the experiments are
indeed done at the lower limit of stability \cite{hofer01b}.
\begin{itemize}
\item
In general the measured magnetic contrast depends on the angle
$\phi_M$ between tip and sample magnetization.
\end{itemize}
Therefore this angle can be uniquely determined by comparing
experimental results with simulations. The experimentally obtained
magnetic contrast was 2 to 4pm \cite{bode01c}. In our simulations
this value indicates either, that the tip was covered by at least
one monolayer of Mn, or that the angle between the two
magnetization vectors was $87-89^{\circ}$ (Fe tip apex or Mn tip
apex). Given that this seems highly improbable for two coplanar
vectors, the most likely explanation of the experiments is that
they were performed with a tip close to our model (c) (see Fig.
\ref{fig001}). However, this tip is the least suitable to resolve
the angle $\phi_M$. With a clean tip, or a tip only contaminated
by single atoms of the sample, this angle could be resolved
accurately to about 1-2$^{\circ}$.

The last step in obtaining all magnetic properties on the atomic
scale is the variation of the angle $\phi_M$. This can essentially
be done in two ways: (i) Either the sample or the STM tip can be
rotated. (ii) The orientation of the magnetic field of the STM tip
is stabilized by an external magnetic field, and this external
magnetic field is rotated. In both cases images have to be
recorded for a number of settings. From these images the direction
in space of the sample magnetization can be uniquely determined.

The method has a number of important implications. First, it can
lead to a detailed understanding, how magnetism depends on the
layer structure of a magnetic multilayer. As the controlled growth
of single layers is experimentally feasible, magnetic properties
can be directly studied, which change with the number of layers,
e.g. the orientation of the magnetization axis \cite{weller95}.
Second, it is possible to study the influence of impurities on the
local magnetic field. Given that magnetic properties have such a
wide range of applications in current technology, this possibility
to study the site-dependency of magnetism on the atomic scale
could lead to a much better understanding of the chemical
determinants of magnetic systems. We think that this potential
alone makes the method, and the SP STM, the most promising tool in
nanomagnetic research.

In summary we have shown that the combination of high resolution
spin-polarized STM scans and first principles simulations of the
tunnel current makes it possible to determine all magnetic
properties of a surface on the atomic scale: the local magnetic
moment as well as the polarization and the direction of surface
magnetization. We expect that this method will ultimately allow to
study changes of the magnetic surface properties in the vicinity
of dopants and impurities.

The work was supported by the British Council and the National
Research Council. Computing facilities at the UCL HiPerSPACE
center were funded by the Higher Education Funding Council for
England.

\end{document}